\documentclass[a4paper]{VisionStyle}
\usepackage{epsfig}

\begin{document}

\title{XMM-Newton Measurements of the Elemental Abundances in the Intracluster Medium}

\author{T.\,Tamura\inst{1}  \and J.S.\,Kaastra\inst{1}  \and  J.A.M.\,Bleeker\inst{1} \and C.\,Ferrigno\inst{1} 
\and J.R.\, Peterson\inst{2}} 

\institute{
SRON National Institute for Space Research, 
Sorbonnelaan 2, 3584 CA Utrecht, The Nether\-lands 
\and 
Astrophysics Laboratory, Columbia University, 
550 West 120th Street, New York, NY 10027, USA
}

\maketitle 

\begin{abstract}
Based on {\it XMM-Newton} observations of a sample of galaxy clusters,
we have measured the elemental abundances (mainly O, Si, S, and Fe) and their spatial distributions in the intracluster medium (ICM).
In the outer region of the ICM,
observations of the O:Si:S:Fe ratio are consistent with the solar value, 
suggesting that the metals in the ICM were produced by a mix of supernovae (SNe) Ia and II.
On the other hand, around the cD galaxy, 
the O/Fe ratios are about half of the solar value because of a central excess of the Fe abundance.
An increase of the relative contribution from SNe Ia in the cD galaxy to the metal production towards the center 
is the most likely explanation.
\end{abstract}

\section{Introduction}
Clusters of galaxies are filled with an X-ray emitting ICM.
The ICM is not only a dominant baryon component in the nearby Universe
but also contains comparable or much amount of heavy elements than those in galaxies.
Therefore, the distribution and composition of metals in the ICM is essential 
for understanding the history of metals in galaxies and clusters.

Following early measurements, 
{\it ASCA} and {\it BeppoSAX} observations have revealed several important properties of the ICM metallicity.
These include Fe abundance increases around cD galaxies (e.g. Fukazawa 1998; De Grandi and Molendi 2001) 
and variations in Si/Fe ratio within a cluster (e.g. Finoguenov et al. 2000) and among clusters (e.g. \cite{ttamura-B3:fu98}).
{\it XMM-Newton} observations with higher capability
should improve the accuracy of the measurements.
In particular, thanks to the high spectral resolution (RGS) and better sensitivity (EPIC) 
in the soft X-ray band, 
significant improvement of the O abundance determination is expected.
The O abundance is crucial to understand the origin of the ICM metal 
since O is expected to be produced mostly by SNe II.

Early results from {\it XMM-Newton} of S\'ersic~159-03, A~1835, A~1795, and A~496 were reported in Kaastra et al. (2001), Peterson et al. (2001), Tamura et al. (2001a) and \cite*{ttamura-B3:ta01b}, respectively.
Here we present new results of abundance measurements.
Table~\ref{ttamura-B3_tbl:sample} summaries our sample.

Note that most of our sample have a giant elliptical galaxy (cD galaxy) at the X-ray center.
Around the galaxy, we often find metal-rich and cooler X-ray emission compared to the outer region of the cluster.

\begin{table}
\caption[]{
The sample. 
(2) redshift. 
(3) The ICM temperature in keV.
(4) {\it XMM-Newton} publications and notes.}
\label{ttamura-B3_tbl:sample}
\begin{center}
    \leavevmode
    \footnotesize
\begin{tabular}{lccl}
\hline
(1) target	& (2) z	& (3) T	& (4) \\
\hline
M87/Virgo	&  0.0044	& 2.5	& BBK01, BSB01\\
NGC~533		&  0.017	& 1.2	& a galaxy group.\\
A~262		&  0.016	& 2.2	& \\
S\'ersic~159-03	&  0.056	& 2.5	& K01\\
A~496		&  0.033	& 4.5	& T01b\\
Hyd-A		&  0.054	& 4.0	& \\
A~1795		&  0.063	& 6	& T01a\\
\hline
\end{tabular}
\end{center}
\end{table}

\section{The EPIC results - The abundance of the ICM}
We used the spatially-resolved EPIC spectra to measure the radial distribution of the ICM properties.
\begin{itemize}
\item
For each cluster,
we extracted the spectra in several annuli around the emission center 
and fitted the MOS1, MOS2, and PN spectra simultaneously 
with a single temperature plasma (1T) model modified by photoelectric absorption.
\item
In general, the single temperature model provides an acceptable fit to each spectrum.
However, at the very center in some clusters (e.g. A~262), an additional cool component is suggested.
\item
In most clusters, 
the temperature continuously decreases towards the center beyond a certain radius.
On the other hand, the iron metallicity increase towards the center.
\item
We confirmed that the MOS and PN provide values of the O, Si, S, and Fe abundances and temperature statistically consistent with each other.
Based on this, we suppose that the uncertainty of the instrument responses 
affect insignificantly on the results.
\item
New results of the fitting are shown in Table~\ref{ttamura-B3_tbl:epic-fits}.
\end{itemize}

\begin{table*}
\caption[]{The PN+MOS fit results of the 1T model. 
(2) The extracted region in radius in arcsec.
(3) The column density in units of $10^{24}$~m$^{-2}$.
(4) The temperature in keV.
(5)-(10) The abundances relative to the solar value of Anders and Grevesse (1989).
For MOS and PN the 0.4--8~keV and 0.5--8~keV bands, respectively, are used. 
The errors in parameters are not shown, except for NGC~533, 
but those are similar level to those of NGC~533.
}
\label{ttamura-B3_tbl:epic-fits}
    \leavevmode
    \footnotesize
\begin{center}
\begin{tabular}{lcccccccccc}
\hline
(1) target& (2)   & (3) $N_\mathrm{ H}$ & (4) T& (5) O & (6) Ne=Mg   & (7) Si & (8) S & (9) Ar=Ca & (10) Fe & $\chi^2/\nu$ \\
\hline
Hyd-A	& 48--128 & 3.6	& 3.5		& 0.31 & 0.47 & 0.33 & 0.37 & 0.32& 0.33 & 527/468\\
Hyd-A	& 128--384 & 2.8	& 4.5		& 0.24 & 0.46 & 0.16 & 0.34 & 0	  & 0.29 & 533/468\\	
A~262	& 128--384 & 7.9	& 2.2		& 0.32 & 0.16 & 0.44 & 0.40 & 0.45 & 0.37 & 503/468\\
NGC~533	& 48--128  & 6.9	& 1.2		& 0.23 & 0    & 0.37 & 0.40 & 0.2   & 0.35 & 383/312\\
	&		& 
($^{+0.10}_{-0.13}$)	& 
($^{+0.06}_{-0.01}$)	& 
($^{+0.14}_{-0.13}$) 	& 
($<0.05		$)	& 
($^{+0.13}_{-0.08}$)	& 
($^{+0.18}_{-0.16}$) 	& 
($<0.48		$)	& 
($^{+0.05}_{-0.04}$) 	& 
			\\
\hline
\end{tabular}
\end{center}
\end{table*}

\section{The RGS results}
\begin{figure*}
\begin{center}
\resizebox{0.32\hsize}{!}{\includegraphics[angle=-90]{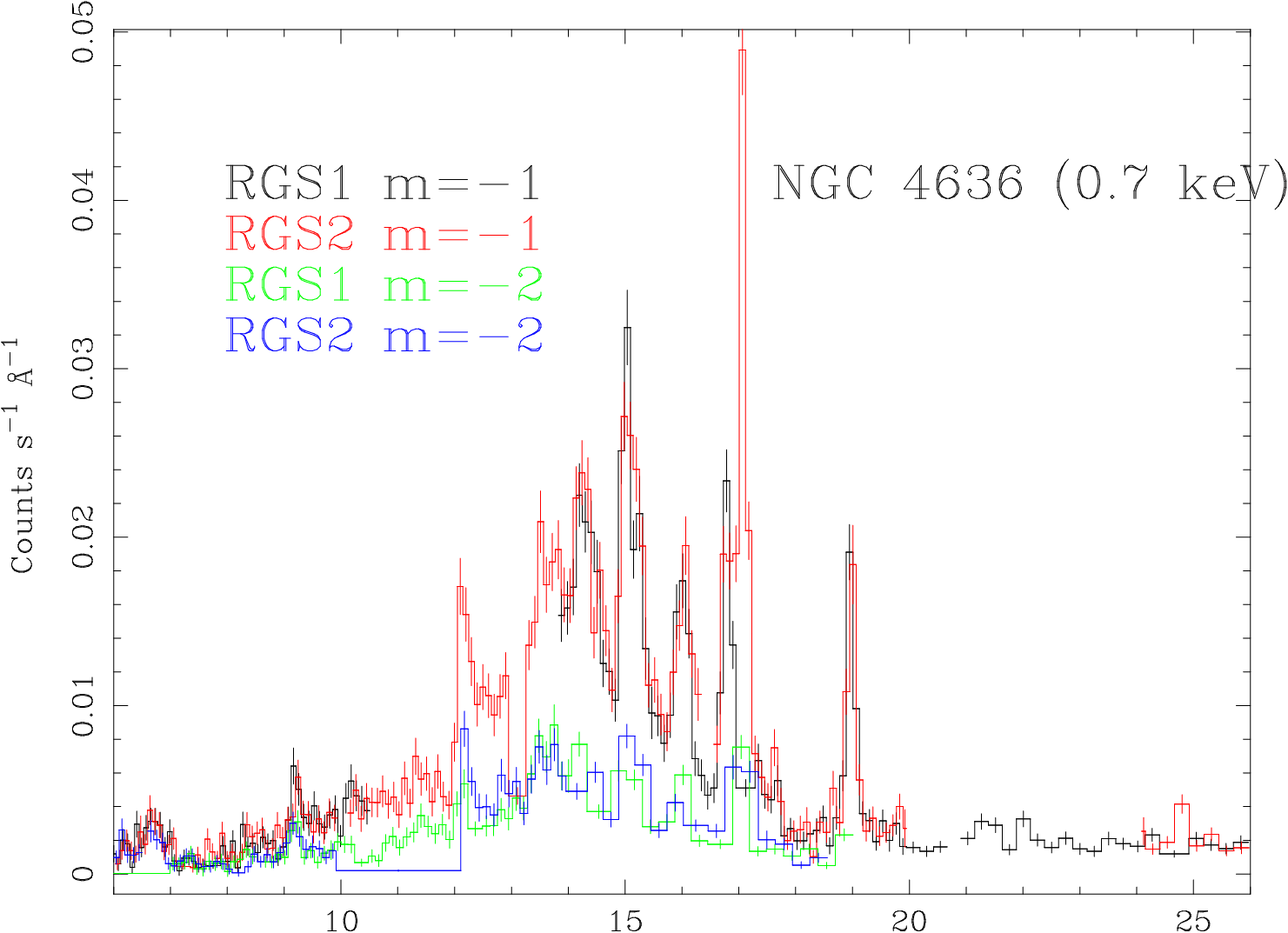}}
\resizebox{0.32\hsize}{!}{\includegraphics[angle=-90]{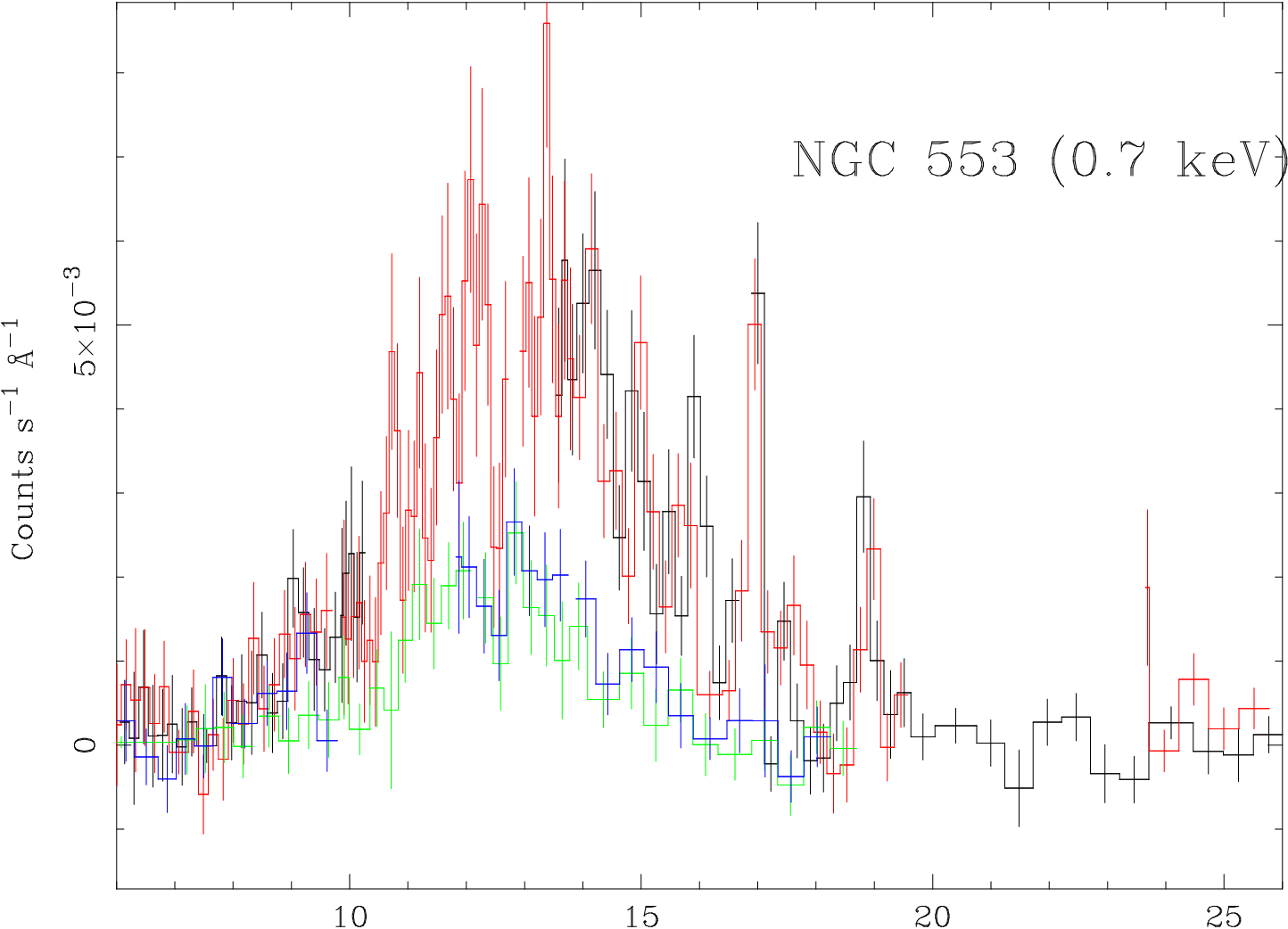}}
\resizebox{0.32\hsize}{!}{\includegraphics[angle=-90]{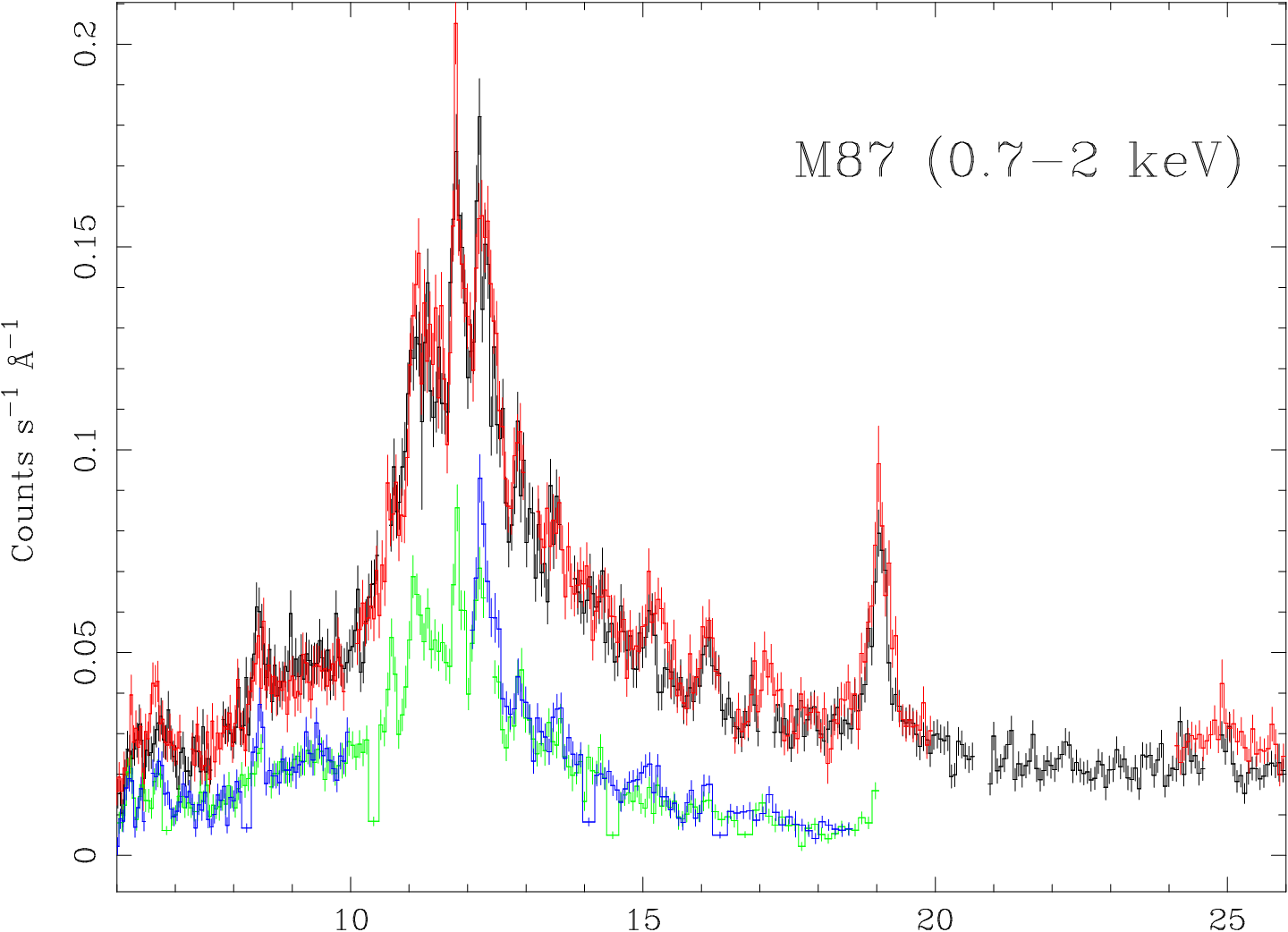}}
\resizebox{0.32\hsize}{!}{\includegraphics[angle=-90]{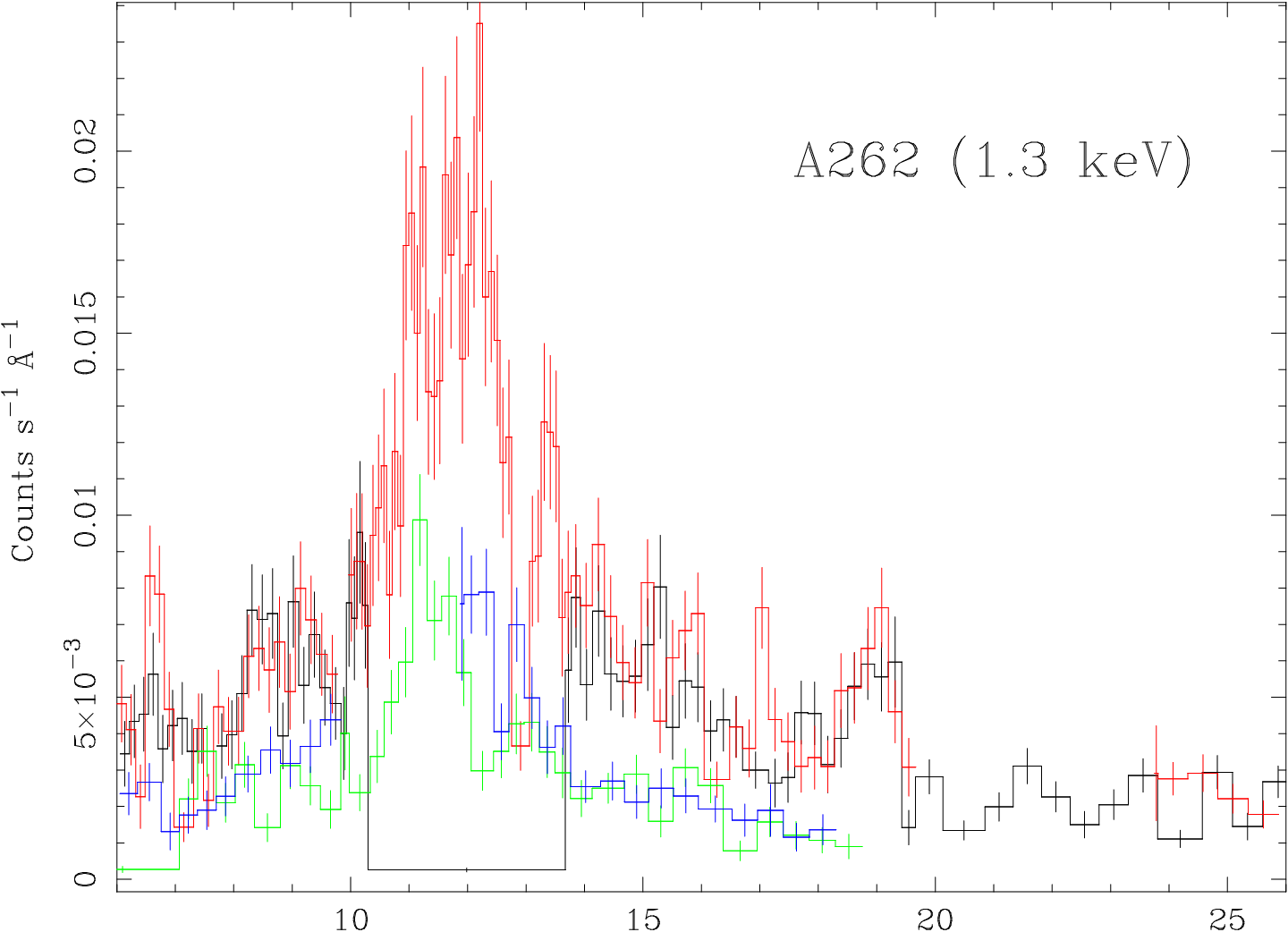}}
\resizebox{0.32\hsize}{!}{\includegraphics[angle=-90]{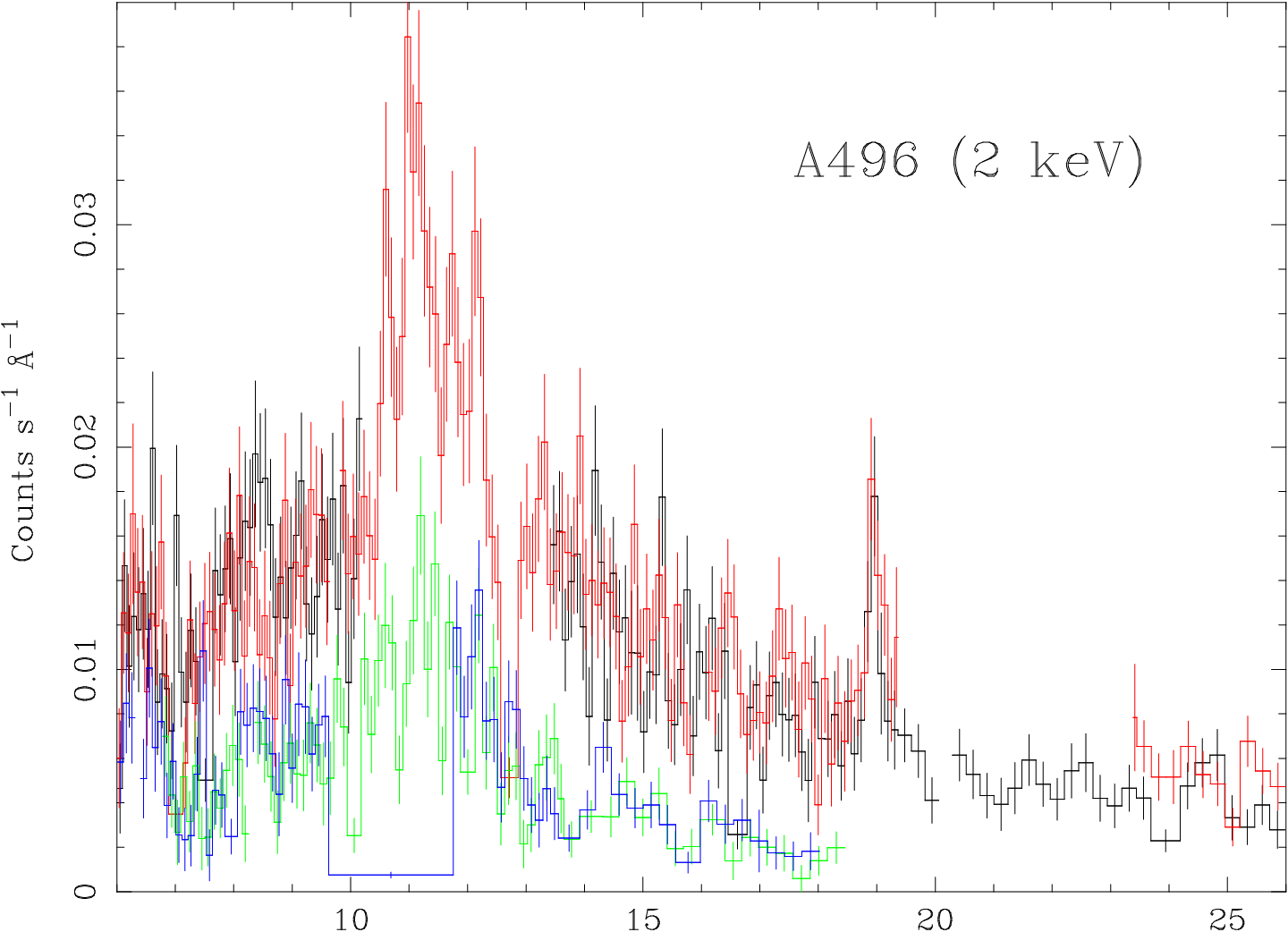}}
\resizebox{0.32\hsize}{!}{\includegraphics[angle=-90]{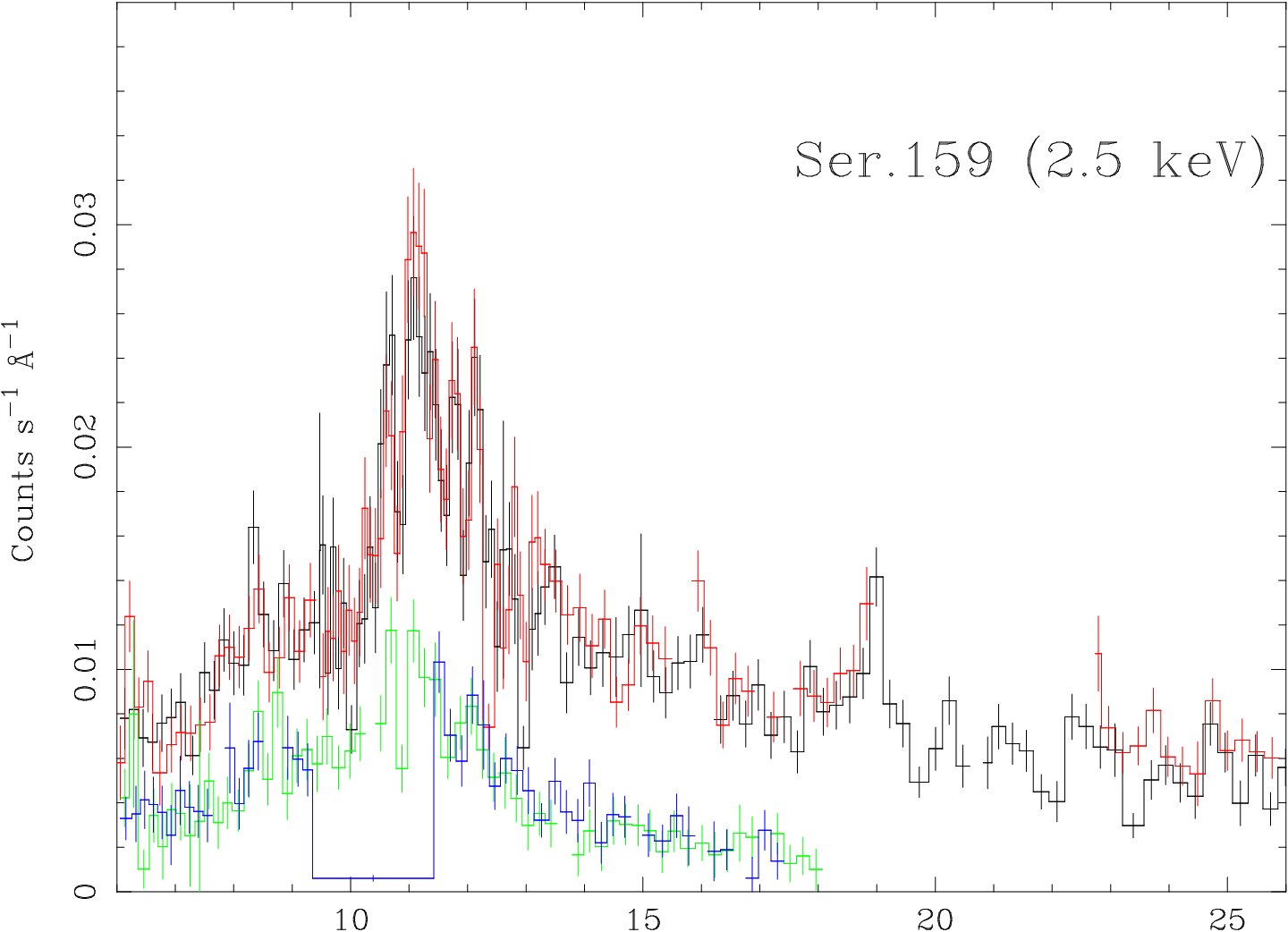}}
\resizebox{0.32\hsize}{!}{\includegraphics[angle=-90]{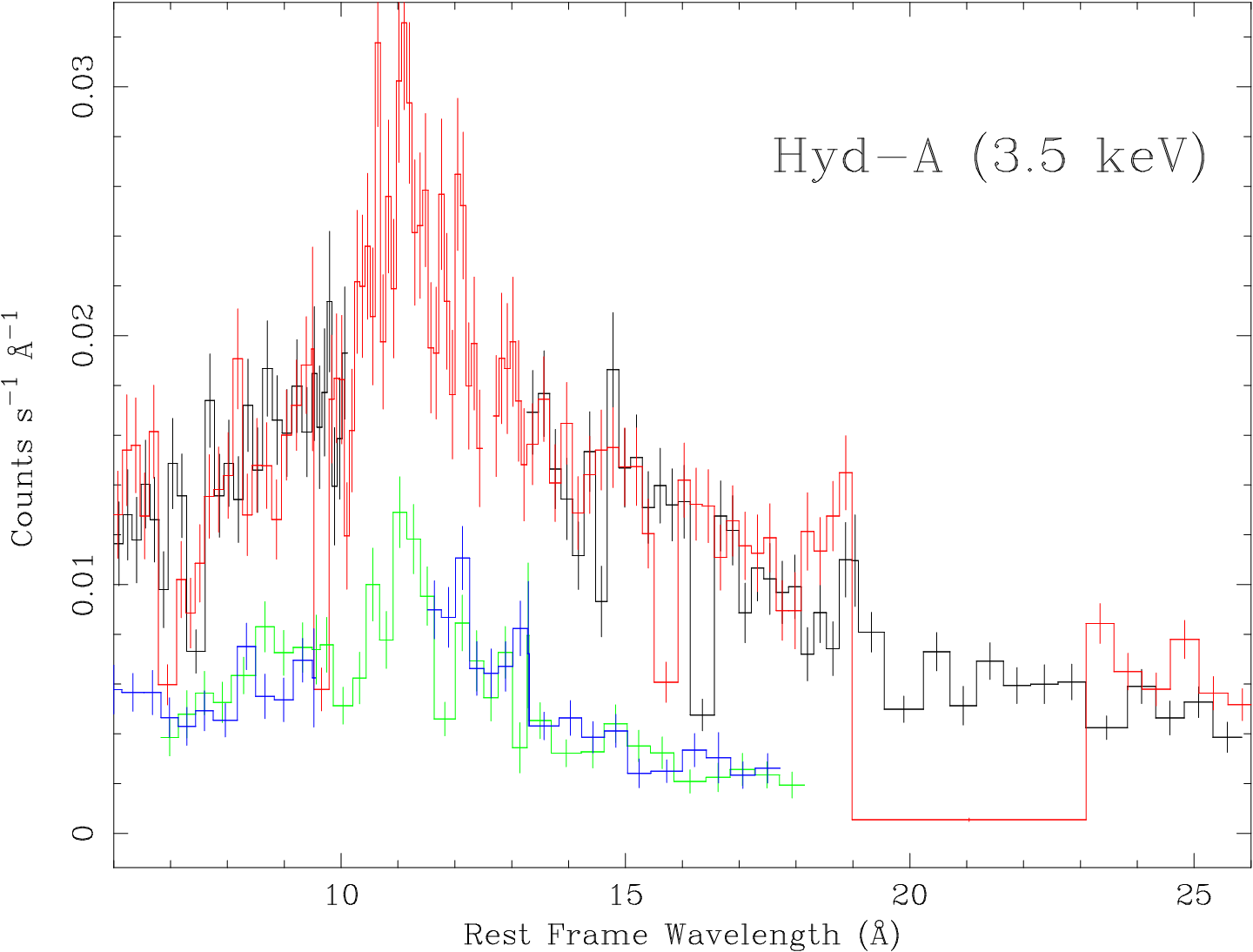}}
\resizebox{0.32\hsize}{!}{\includegraphics[angle=-90]{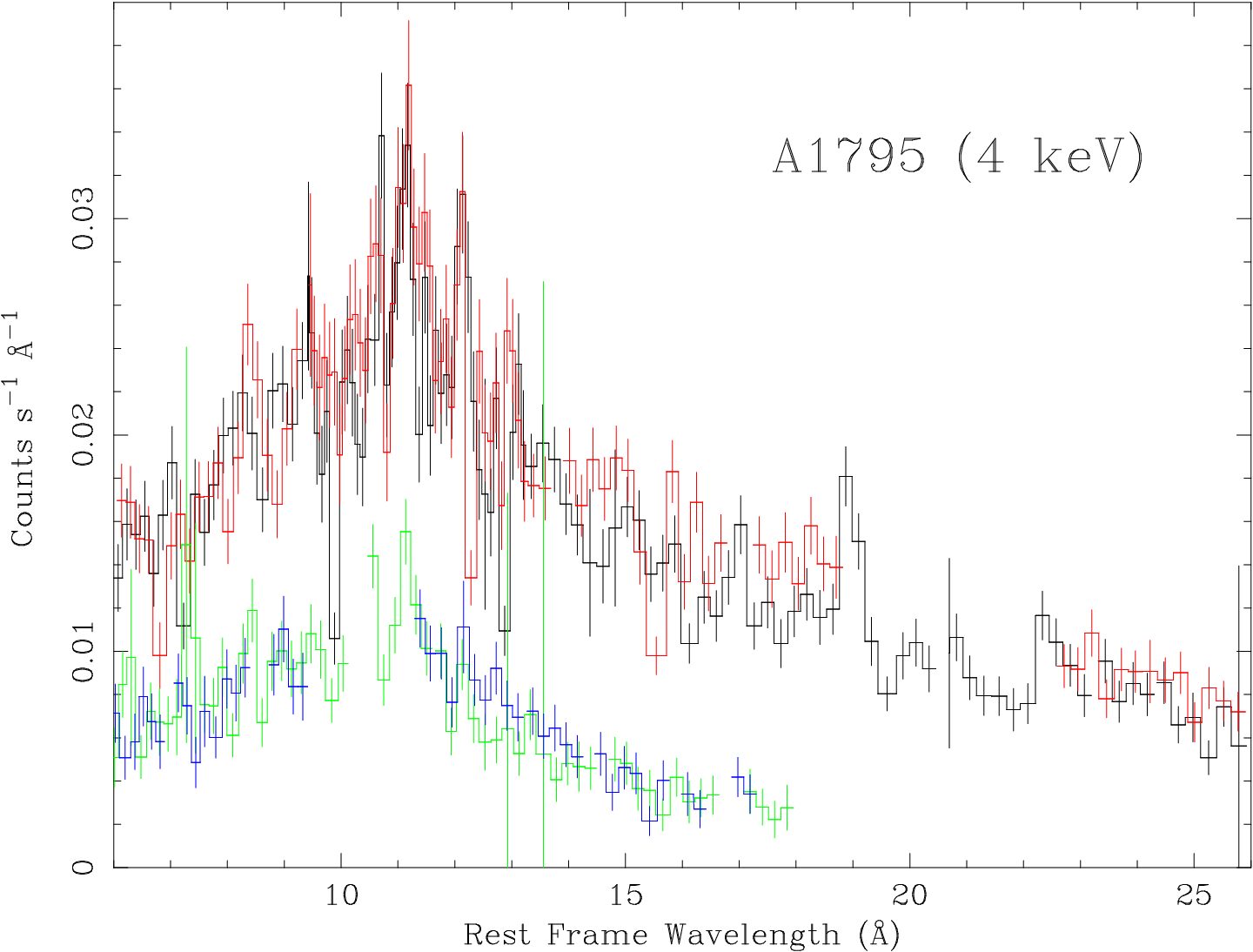}}
\resizebox{0.32\hsize}{!}{\includegraphics[angle=-90]{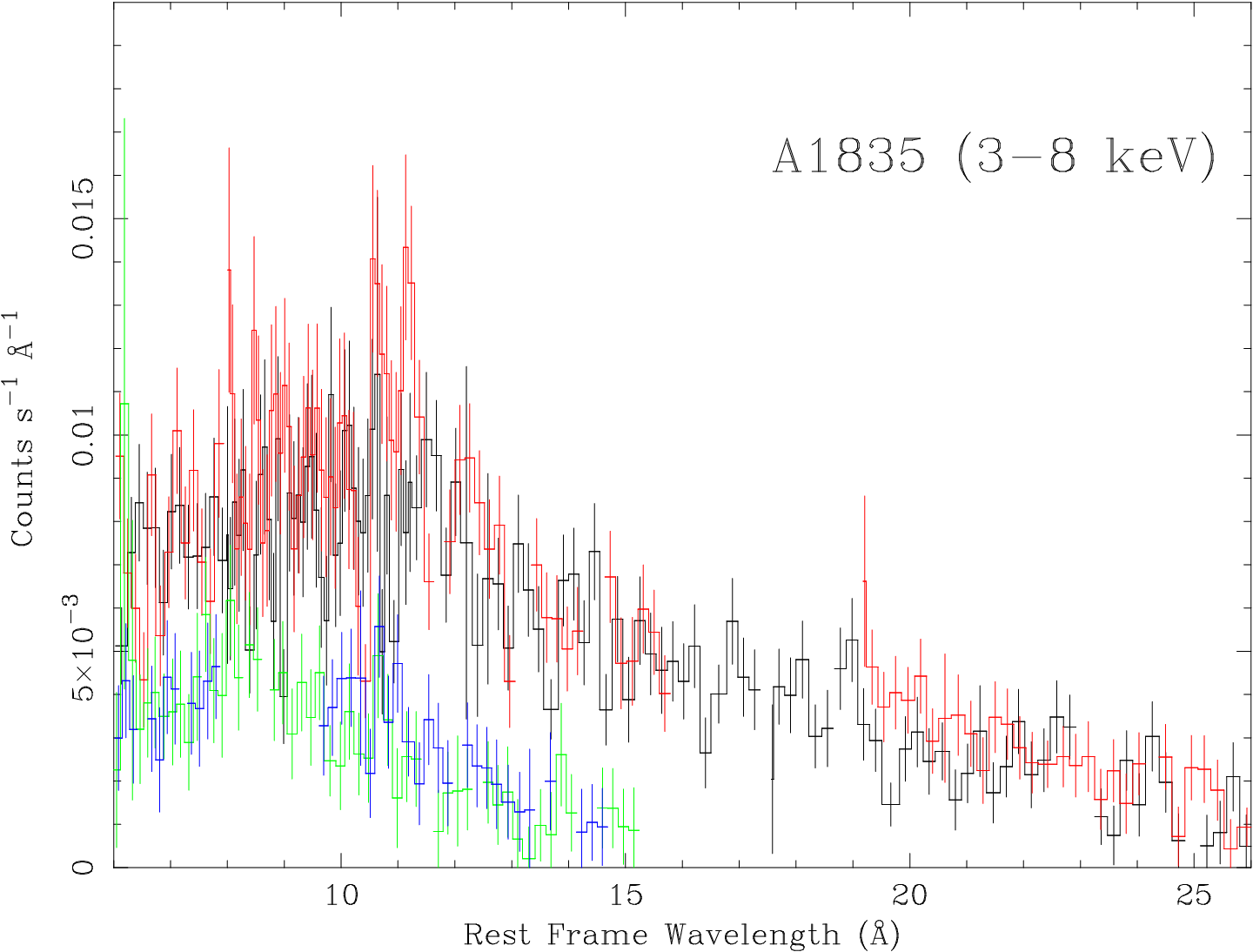}}
\caption{The RGS spectra of cluster centers. No correction for the instrumental responses was made. 
For comparisons, the spectra of a elliptical galaxy, NGC~4636, is also shown (see Xu et al. 2002).
Along with the target name, estimated temperature are indicated.
To help line identification, the CIE model spectra are shown in the next figure.
}
\label{ttamura-B3_fig:rgs-spec}
\end{center}
\end{figure*}

We used the RGS data to constrain the temperature structure and the O and Fe abundance around the cD galaxy at the cluster core.
\begin{itemize}
\item
The spectra of the cluster core were extracted within 
a $\sim 1'$ of full width of the cross dispersion position (Fig.~\ref{ttamura-B3_fig:rgs-spec}).
\item
We calculated the line spread function based upon the source surface brightness profile from the MOS image
and convolved this with the RGS response for a point source.
\item
The 1T or two temperature (2T) models were used. The results are shown in Tables~\ref{ttamura-B3_tbl:rgs-fits} and \ref{ttamura-B3_tbl:rgs-fits-2t}.
\item
In A~262 and NGC~533, the O and Fe abundance depend on the thermal model.
\end{itemize}

\begin{table*}
\caption[]{RGS fit results of 1T model. 
(3) The volume emission measure in units of $h_{50}^{-2}10^{72}$m$^{-3}$.
See the caption in the previous table for other parameters.
Parameters with f were held fixed.
}
\label{ttamura-B3_tbl:rgs-fits}
    \leavevmode
    \footnotesize
\begin{center}
\begin{tabular}{lcccccc}
\hline
(1) target	& (2) $N_\mathrm{ H}$	& (3) EM & (4) T	& O	& Fe	& $\chi^2/\nu$\\
\hline
NGC~533		& 24			& 0.35	& 0.7	& 0.25 ($^{+0.26}_{-0.13}$) & 0.21 ($^{+0.14}_{-0.07}$) 	& 86/125\\
A~262		& 13			& 1.9	& 1.3	& 0.20	& 0.23	& 234/243\\
A~496		& 10 ($^{+1.5}_{-1.5}$) & 5.2	& 2.2 ($^{+0.4}_{-0.3}$)	& 0.23 ($^{+0.08}_{-0.07}$) 	& 0.43 ($^{+0.33}_{-0.25}$) & 314/358\\
S\'ersic~159	& 3.2			& 17	& 2.7	& 0.29	& 0.55	& 360/343\\
Hyd-A		& 7.5			& 24	& 3.4	& 0.23	& 0.41	& 291/292\\
A~1795		& 5.3			& 39	& 4.0f	& 0.31	& 0.44	& 372/346\\
\hline
\end{tabular}
\end{center}
\end{table*}

\begin{table*}
\begin{center}
\caption[]{RGS fit results of 2T model. EM1 (EM2) and T1 (T2) are the emission measure and temperature of the first (second) component, 
respectively.}
\label{ttamura-B3_tbl:rgs-fits-2t}
    \leavevmode
    \footnotesize
\begin{tabular}{lcccccccc}
\hline
target		& $N_\mathrm{ H}$ & EM1	& T1	& EM2	& T2	& O	& Fe	& $\chi^2/\nu$\\
\hline
NGC~533		& 12	& 0.07	& 0.7	& 0.06	& 1.2f	& 0.4 & 0.6 & 72/124\\
A~262		& 8	& 0.7	& 1.0	& 0.5	& 2.5f	& 0.5 & 1.0 & 210/242\\
A~496		& 9.0	& 0.16  & 1.0f	& 4.4	& 2.6	& 0.29 & 0.68 & 303/357\\ 
\hline
\end{tabular}
\end{center}
\end{table*}

\begin{figure*}
\caption{The CIE model (mekal) spectra with indicated temperatures and 0.5 solar abundance. 
Lines from elements are shown in different colors.}
\begin{center}
\resizebox{0.32\hsize}{!}{\includegraphics[angle=-90]{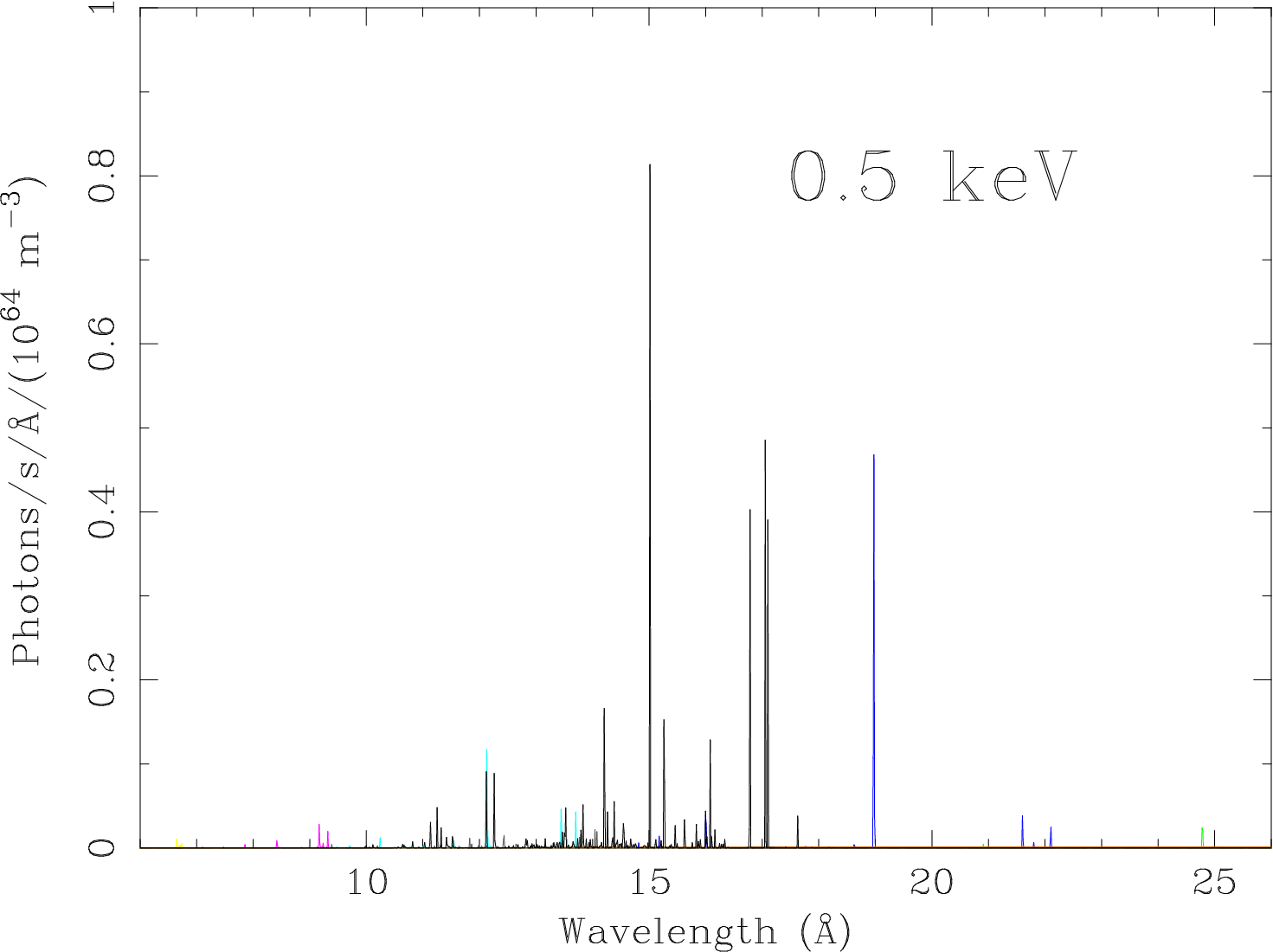}}
\resizebox{0.32\hsize}{!}{\includegraphics[angle=-90]{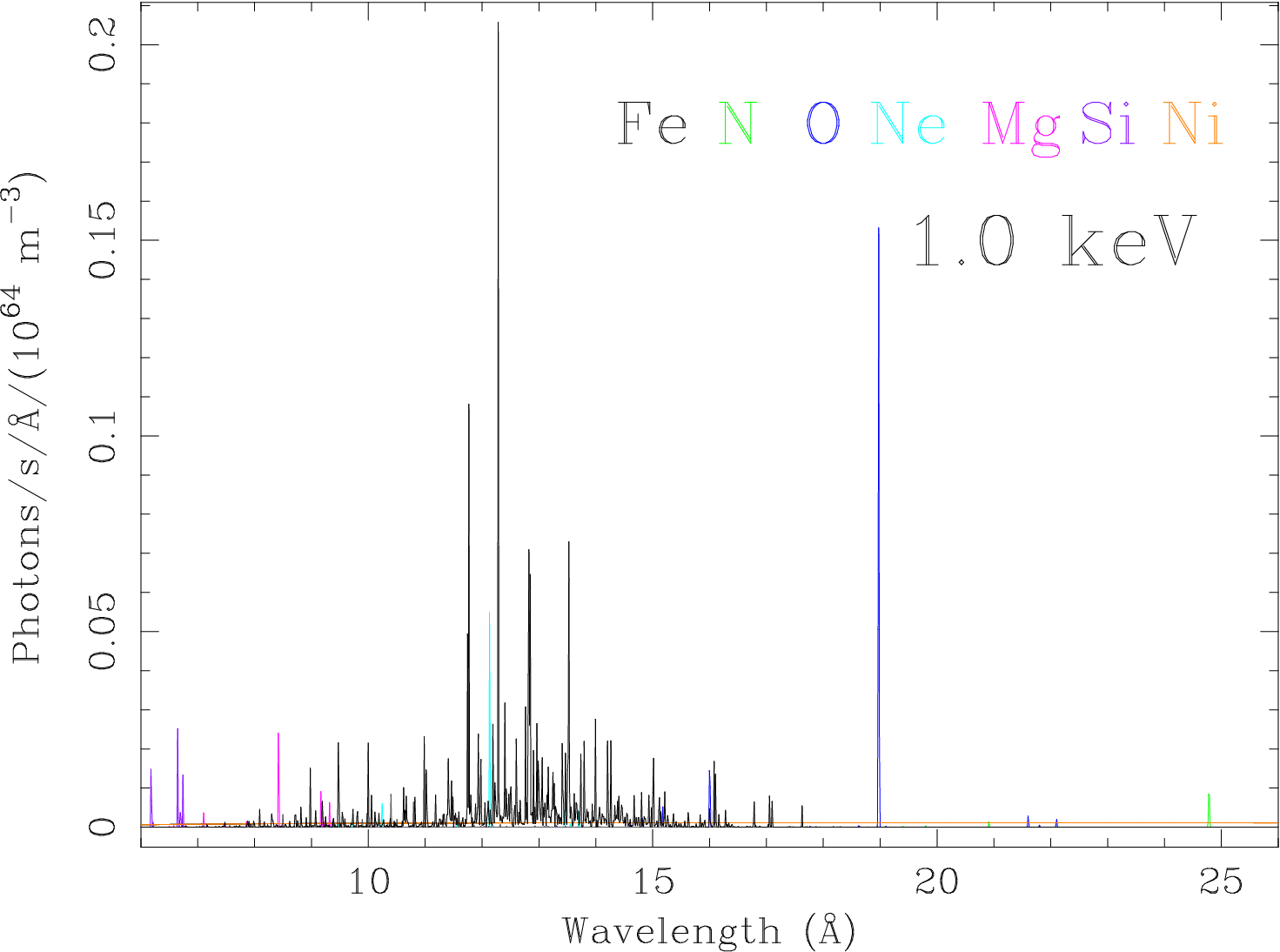}}
\resizebox{0.32\hsize}{!}{\includegraphics[angle=-90]{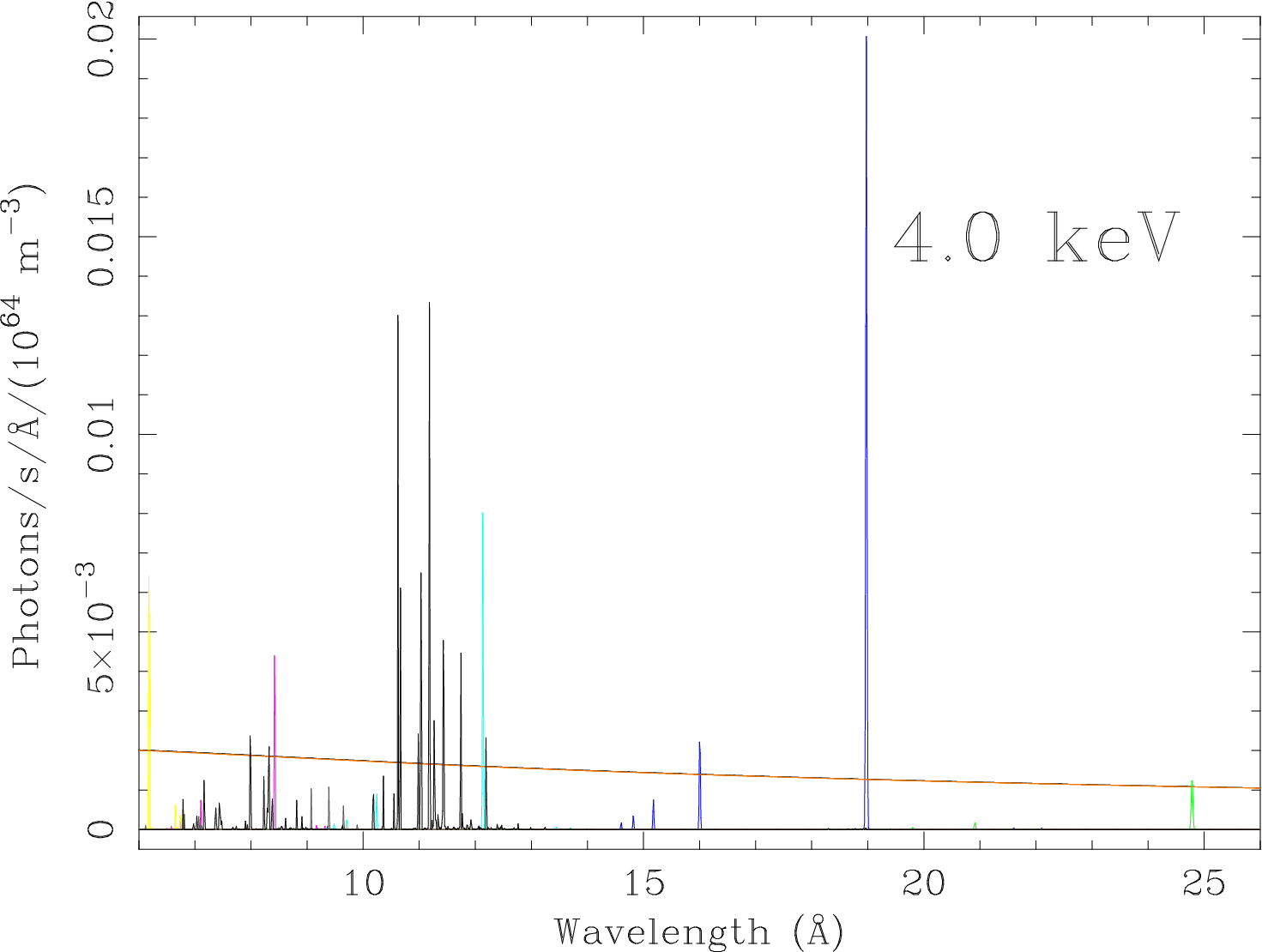}}
\end{center}
\end{figure*}

\section{Summary and Discussion}
\begin{figure}
\resizebox{0.8\hsize}{!}{\includegraphics[angle=-90]{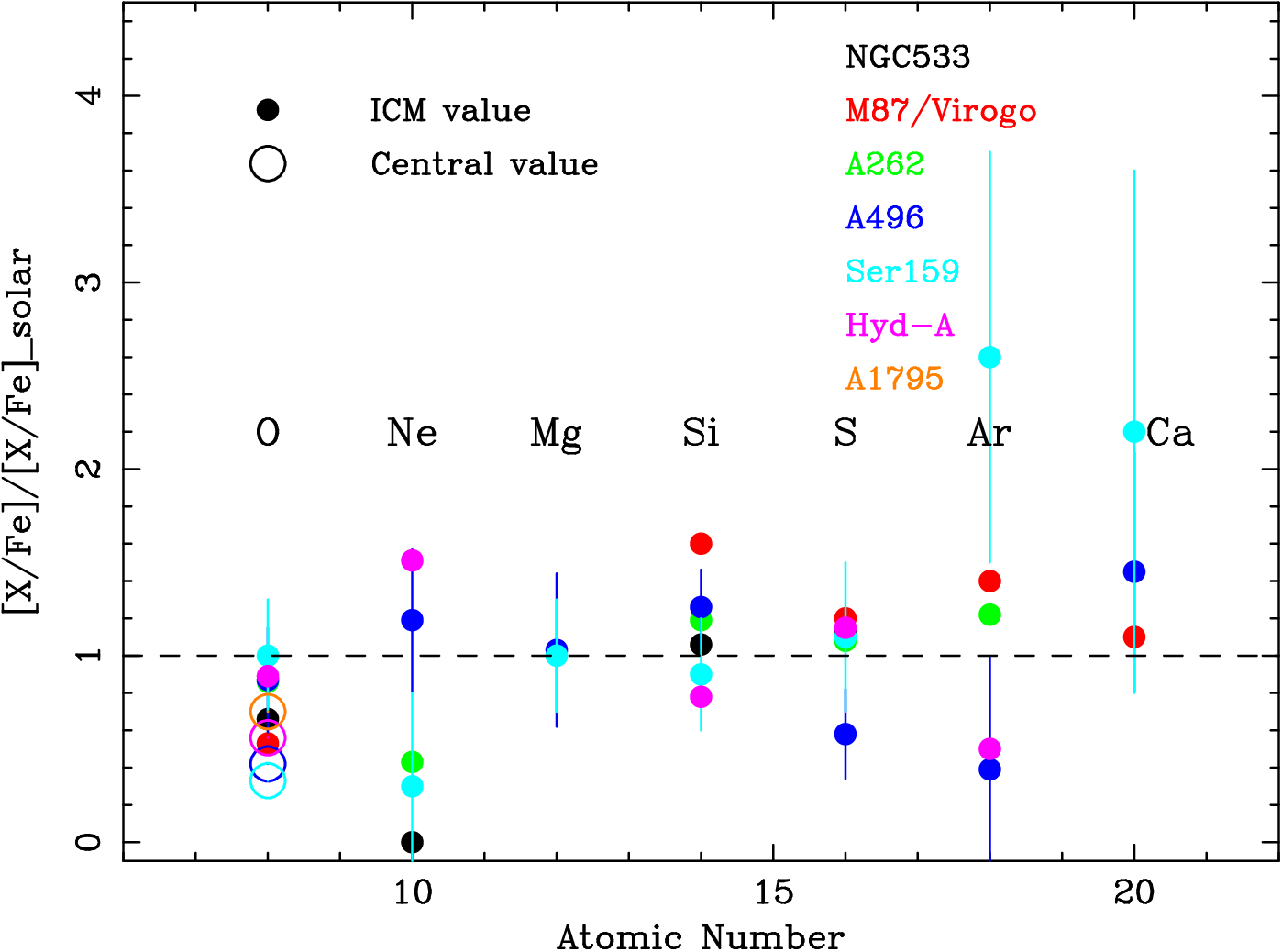}}
\caption{The observed abundance relative to Fe normalized to the solar value.
The values of the central cool component (from RGS) and the ICM (from EPIC) are indicated by a open circle and a filled-circle, respectively.
}
\label{ttamura-B3_fig:abundance}
\end{figure}

Our measurements of the elemental abundances along with the {\it XMM-Newton} results of M87 (B\"ohringer et al. 2001) are summaried in Fig.~\ref{ttamura-B3_fig:abundance}. Main results are followings.
\begin{enumerate} 
\item
The average (and standard variance among clusters) of the O/Fe ratio around the cD galaxy
are 0.5 (and 0.16) times the solar value, respectively.
\item
Those of the O/Fe, Ne/Fe, Si/Fe, S/Fe, and (Ar=Ca)/Fe in the outer region of clusters 
are 0.80 (0.17), 0.7 (0.6), 1.11 (0.29), 1.06 (0.24), and 1.2 (1.0) solar, respectively. 
These variances among clusters are comparable to the statistical errors and insignificant.
\item
There is significant change in the O/Fe ratio between the center and outer regions of clusters.
\end{enumerate}

\subsection{Comparisons with {\it ASCA} results}
Fukazawa et al.(1998) measured Si and Fe abundances in $\sim$ 40~clusters excluding the central cool region.
They reported that the Si/Fe ratio varies depending on the ICM temperature;
groups and poor clusters ($1\sim2$~keV) have Si/Fe of $\sim$ 1 solar, 
while rich clusters exhibit a higher value of 2--3 solar.
Since our sample consists of groups and poor clusters (1.2--4~keV), 
our results of Si/Fe $\sim$ 1 solar is roughly consistent with the ASCA measurements.

\subsection{Origin of the metals in the ICM}
We interpret our new abundance measurements along with other {\it XMM-Newton} results.
Here we assume that the distribution of the abundance ratios in the ICM
have not change since the ejection of those metals from galaxies.
However, we should note that the change in Si/Fe ratio among clusters from the ASCA results suggests that 
a large amount of $\alpha$-elements such as O, Si, and S have selectively escaped from the gravitational potential in poorer systems (Fukazawa et al. 1998).

\begin{figure*}
\begin{center}
\resizebox{0.30\hsize}{!}{\includegraphics[]{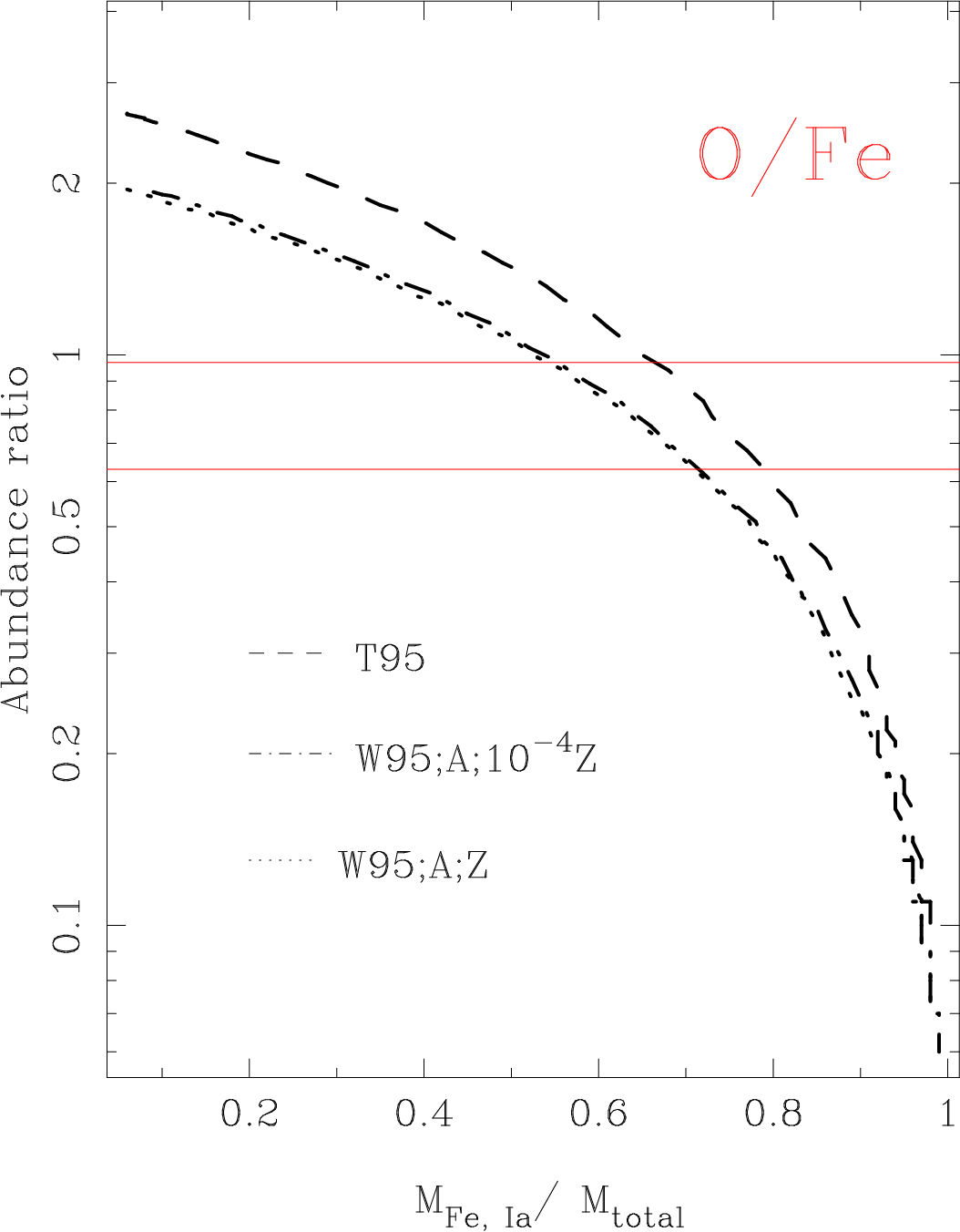}}
\resizebox{0.30\hsize}{!}{\includegraphics[]{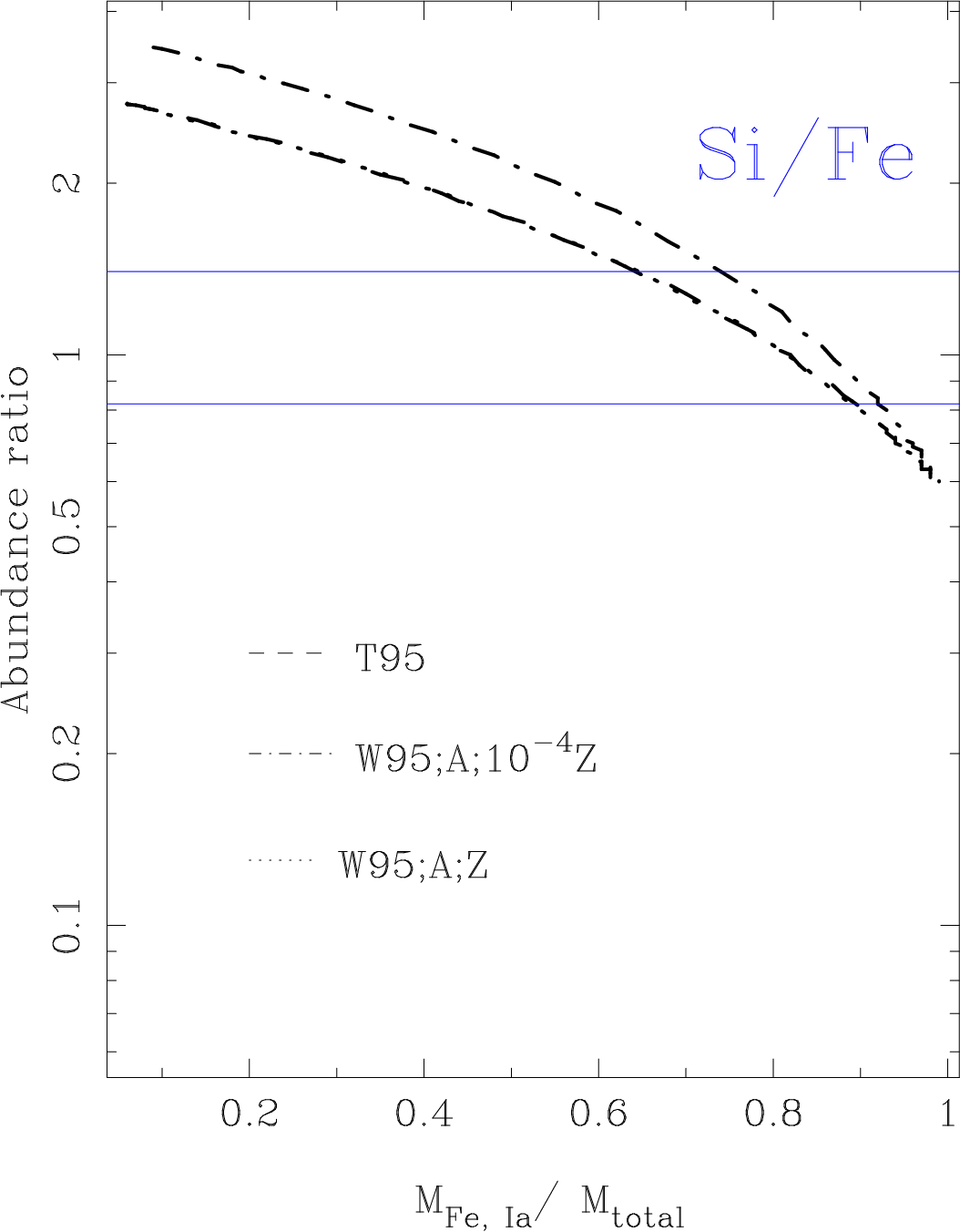}}
\resizebox{0.30\hsize}{!}{\includegraphics[]{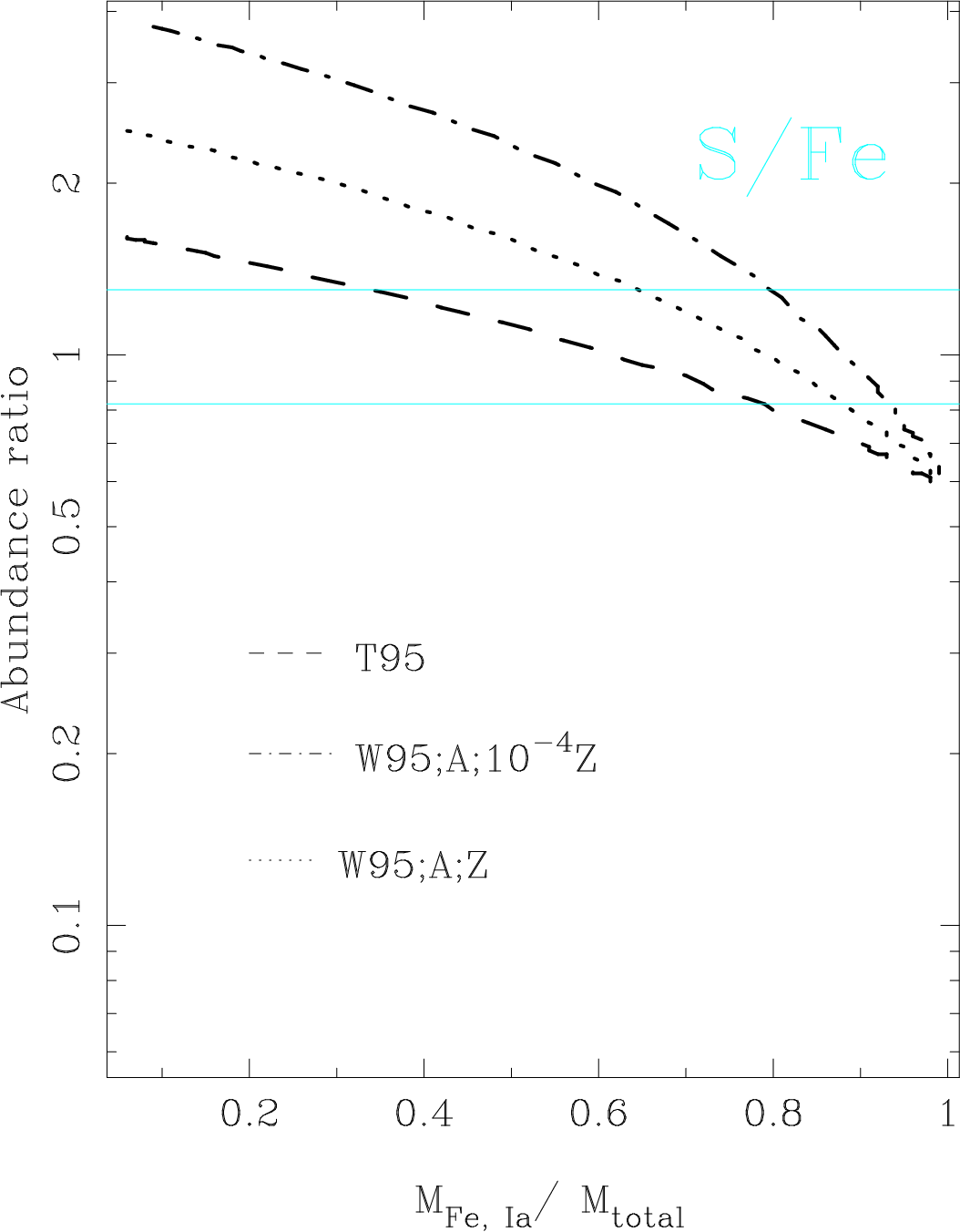}}
\caption{The O/Fe, Si/Fe, and S/Fe ratios: Observation vs. SN Ia$+$SN II model predictions.
Our results of the abundance range are shown in horizontal lines.
As a function of the ICM SN Ia originated fraction of Fe, 
predictions based on a SN Ia yield (Thielemann et al. 1993) and SN II yields from three different models are plotted .
Three SN II models are T95=Tsujimoto et al. (1995), 
W95;A;Z and W95;A;$10^{-4}$ = model A in Woosley \& Weaver (1995) with Z=Z$_{\odot}$ and Z=$10^{-4}$Z$_{\odot}$, where Z is metallicity of the stars, respectively. 
We used yields calculations by Gibson et al. (1997) who averaged elemental yields over the progenitor mass range 10--50~\hbox{M$_{\odot}$} for a Salpeter IMF.
}
\label{ttamura-B3_fig:sniaii}
\end{center}
\end{figure*}

The result (2)
indicates that 
in member galaxies in the clusters metal has been enriched in a similar way to that in solar neighborhood,
and then ejected into the intergalactic space.
Then, we compare the observed abundance ratios among O, Si, S, and Fe with a mixture of SN Ia and SN II prediction in Fig.~\ref{ttamura-B3_fig:sniaii}.
This comparison indicates that the observed O:Si:S:Fe ratio is consistent with a SN Ia originating Fe mass fraction of 0.6--0.8 (or equivalently a SN Ia/SN II frequency ratio of 0.2--0.7).

How about the cluster center ?
The result (3) implies that 
there are at least two different origins for the metals in the ICM,
irrespective of any theoretical model for the metal production.
An increase of the relative contribution from SNe Ia in the cD galaxy to the metal production towards the center is one possibility.
This is because a SN Ia is supposed to produce O/Fe ratio smaller than the solar value.
In fact, Fukazawa et al. (1998) found that the excess Fe mass around the cD galaxy in general can be produced by the standard SN Ia rate over a Hubble time.
The abundances of other elements such as Si and S at the cluster center, 
which is not addressed here, 
are important to examine this idea more quantitatively (Fukazawa 1998; Finoguenov et al. 2000).

\end{document}